\documentclass[prl, twocolumn, superscriptaddress, floatfix,, amsmath,amssymb, aps]{revtex4-2}
\usepackage{graphicx}
\usepackage{dcolumn}
\usepackage{bm}
\usepackage{hyperref}
\usepackage{siunitx}
\DeclareSIUnit\gauss{G}
\usepackage{braket}
\usepackage{ulem}
\usepackage[dvipsnames]{xcolor}
\usepackage{tabularx}
\usepackage{braket}

\begin{document}

\title{Asymmetric trions in monolayer transition metal dichalcogenides}

\author{Arthur Christianen}
\affiliation{Institute for Theoretical Physics, ETH Zürich, Zürich, Switzerland}
\affiliation{Institute for Quantum Electronics, ETH Zürich, Zürich, Switzerland}

\author{Ata\c{c} {\.I}mamo{\u{g}}lu}
\affiliation{Institute for Quantum Electronics, ETH Zürich, Zürich, Switzerland}

\date{\today}

\begin{abstract}
Exciton spectroscopy serves as a sensitive probe of electronic states in two-dimensional semiconductors. A prominent feature in optical spectra is the trion peak arising from the binding of a charge carrier to an exciton. The splitting between the exciton and trion peaks is usually interpreted as the trion binding energy, but we theoretically show that this view is incomplete. Since dark excitons are more strongly bound than the bright exciton, the trion wave function is asymmetric and a large contribution to the measured splitting is the difference between the bright and dark exciton binding energies. Our model quantitatively explains the measured trion energies in MoSe$_2$ and WSe$_2$, demonstrating the importance of the internal structure of the exciton for the interpretation of the optical response of transition metal dichalcogenides.
\end{abstract}

\maketitle

Two-dimensional semiconductors, particularly transition metal dichalcogenides (TMDs), exhibit a rich variety of correlated and topological phases. In monolayer TMDs, Wigner crystal formation has been reported \cite{smolenski:2021}, while twisted bilayer configurations host more exotic states, including fractional Chern insulators \cite{cai:2023,xu:2023,zeng:2023} and superconductivity \cite{xia:2025,guo:2025}. 

Exciton spectroscopy provides a powerful and versatile probe of these phases because the injected excitons interact with the charge carriers, leading to line shifts and new spectral features. This approach therefore enables the quantitative extraction of key parameters such as the carrier density, valley polarization, and the strength of the moiré potential \cite{kiper:2025}. Beyond their utility as probes, exciton–polaron spectra reveal features of intrinsic interest, such as the emergence of large excitonic complexes \cite{vantuan:2022,dijkstra:2025}, which themselves constitute strongly correlated electronic states. 

While most main features of exciton–polaron spectra are qualitatively understood \cite{sidler:2017,efimkin:2017,courtade:2017,glazov:2020,vantuan:2022}, even in monolayer TMDs critical theoretical advances are needed to achieve quantitative agreement with experiment. For example, in WSe$_2$, the binding energy of the trion, a bound state of an exciton with an additional charge carrier, has only matched experimental values \cite{courtade:2017} when the dielectric constant of the surrounding hexagonal boron nitride (hBN) is adjusted away from independently measured values \cite{stier:2018,goryca:2019}. Improving our theoretical understanding is therefore essential to extract reliable information from exciton spectra, especially as the complexity of the samples increases.

\begin{figure}[!b]
    \centering
\includegraphics[width=0.62\columnwidth]{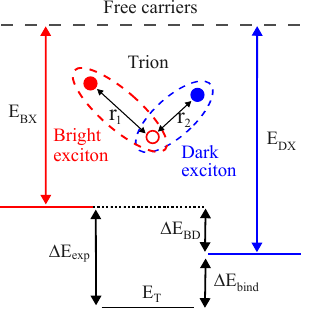}
    \caption{Schematic overview of the trion structure and its binding in TMDs. In TMDs the dark exciton binding energy $E_{\mathrm{DX}}$ is larger than the bright exciton binding energy $E_{\mathrm{BX}}$. This results in an asymmetric trion, with $\langle r_2 \rangle < \langle r_1 \rangle$. Consequently, the trion binding energy $\Delta E_{\mathrm{bind}}$ should be defined with respect to the dark exciton. The experimentally measured splitting $\Delta E_{\mathrm{exp}}$ between the bright exciton and the trion therefore does not only depend on $\Delta E_{\mathrm{bind}}$, but also on $\Delta E_{\mathrm{BD}}=E_{\mathrm{DX}}-E_{\mathrm{BX}}$. See Fig.~\ref{fig:bandstructures}a) for a level diagram including the optical transitions.}
    \label{fig:trion_scheme}
\end{figure}

In this Letter, we show that trion wave functions can be strongly asymmetric—qualitatively different from conventional assumptions. Incorporating this asymmetry, we quantitatively reproduce the experimental trion energies \cite{courtade:2017,kiper:2025} using input from independent measurements \cite{stier:2018,goryca:2019} and density functional theory (DFT) calculations \cite{li:2022}.

The asymmetry of the trion arises from the difference in binding energy between bright and dark excitons, see Fig.~\ref{fig:trion_scheme}a). Due to this splitting, the two same-charge carriers in a trion no longer form pure spin singlet or triplet states with strictly symmetric or antisymmetric spatial wave functions.
In TMDs, dark excitons are more strongly bound than bright excitons \cite{zhang:2017,robert:2020} due to the mass difference between the conduction bands \cite{kormanyos:2015}, and/or the absence of the repulsive exchange interactions present in bright excitons \cite{li:2022}. 

Although the bright–dark exciton binding energy difference $\Delta E_{\mathrm{BD}}$ is small relative to the exciton binding energy, it is comparable to the trion binding energy. As a result, the bond length $r_1$ of the bright exciton within the trion is significantly larger than $r_2$, its dark exciton counterpart. This makes the dark exciton the natural reference state for defining the trion binding energy.

Consequently, the experimentally measured energy difference $\Delta E_{\mathrm{exp}}$  between the trion and bright exciton features does not directly yield the trion binding energy, as it also includes $\Delta E_{\mathrm{BD}}$ [see Fig.~\ref{fig:trion_scheme}a)]. We show that in WSe$_2$ $\Delta E_{\mathrm{BD}}$ is even the dominant contribution to the measured splitting, leading to an important reinterpretation of the exciton spectrum.

In the following, we will first introduce the theoretical approach. Then we will discuss the bright trion binding energies and wave functions.  Finally, we will conclude and give an outlook.

\textit{Theory-}
We consider hBN-encapsulated monolayer WSe$_2$ and MoSe$_2$ in the low-doping regime, where the effect of many-body dressing on the exciton is small. We numerically compute the trion energies using an approach analogous to Ref.~\cite{fey:2020} (see Supplemental Material). We assume a quadratic dispersion for the electrons and holes and express the trion wave function as 
\begin{equation}
    \psi_T(r_1,r_2,\theta, \alpha)=\frac{u(r_1,r_2,\theta)}{\sqrt{2\pi r_1 r_2}} \exp(i m \alpha),
\end{equation}
where $r_1$ and $r_2$ are the electron-hole distances (see Fig.~\ref{fig:trion_scheme}), $\theta$ is the angle between $\vec{r}_1$ and $\vec{r}_2$, and $\alpha$ specifies the global orientation of the trion. We focus on the total angular momentum zero sector ($m=0$). We variationally expand the wave function $u(r_1,r_2,\theta)$ in a basis of 2D-hydrogen wave functions [see Supplemental Material Eqs.~\eqref{eq:varwf1} and \eqref{eq:varwf2}], fully including the $\theta$-dependence.

As usual \cite{stier:2018,goryca:2019,fey:2020}, we model the direct Coulomb interactions in the trion via the Rytova-Keldysh potential \cite{rytova:1967,keldysh:1979}
\begin{equation}
V_{RK}(r)= \frac{\pi}{2 r_0}\left[H_0\left(\frac{\kappa r}{r_0} \right)-Y_0\left( \frac{\kappa r}{r_0}\right) \right].
\end{equation}
Here, $H_0$ and $Y_0$ are the zeroth-order Struve function and Bessel function of the second kind. At large $r$, this potential asymptotically recovers the Coulomb form $1/(\kappa r)$. The screening length $r_0$ captures the polarizability of the monolayer, while $\kappa$ is the effective dielectric constant of the surrounding environment.  Both parameters have been experimentally extracted from fits to Rydberg exciton spectra in hBN-encapsulated TMDs \cite{goryca:2019}.

Several effects lead to deviations from the idealized Rytova-Keldysh potential, such as the finite thickness of the TMD layers and the different in-plane and out-of-plane polarizabilities of hBN. In addition, the anharmonicity and the quantum geometry \cite{srivastava:2015} of the valence and conduction bands \cite{liu:2013,kormanyos:2015} play a role. Since these effects mostly play a role at short distances, we assume that these effects can be approximately described by absorbing them into the $r_0$ parameter. Consequently, the experimentally extracted value of $r_0$ \cite{goryca:2019} should not be viewed as a true microscopic parameter, but as an effective screening parameter that ensures the correct 1s exciton energy. Excited-state energies are less sensitive to its precise value.

In addition to the direct Coulomb interaction, exchange interactions play a crucial role in determining the trion energies, as they give rise to an energy difference between bright and dark excitons. For an exciton with electron-hole separation $r$ and wave function $\psi_X(r)$, the exchange energy $X$ is given by \cite{li:2022}
\begin{equation}
    X = |\psi_X(r=0)|^2 I_X,
\end{equation}
where $I_X$ depends on the Bloch wave functions of the relevant conduction and valence bands and can be estimated using density functional theory \cite{li:2022}. We model the exchange interaction as a short-range Gaussian potential
\begin{equation}
    V_X(r)= c_X \exp(-\frac{r^2}{\sigma_X^2}),
\end{equation}
with small $\sigma_X=2${\AA} and $c_X$ chosen such that the integrated potential recovers $I_X$. We neglect the exchange interactions between opposite-spin electron-hole pairs, as they are estimated to be an order of magnitude smaller than for the same spin case \cite{li:2022}.
In WSe$_2$, bright and valley-dark excitons are predicted to have similar exchange energies (11.5 and 11.4 meV respectively), but they still have a substantial binding energy difference due to the different masses of the upper and lower conduction bands. The value of $I_X$ is higher for bright excitons, but the stronger binding of the dark exciton leads to larger $|\psi(r=0)|^2$, resulting in nearly equal exchange energies. Exchange interactions also exist between same-spin electrons in different valleys \cite{dery:2016}, but their effect on the trion energy is small, since the probability density at zero electron–electron separation, $|\psi(r_{ee}=0)|^2$, is strongly suppressed due to the Coulomb repulsion. We therefore neglect these terms.

\begin{figure}[t!]\includegraphics[width=0.45\textwidth]{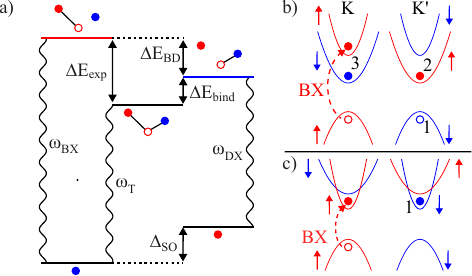}
    \caption{a) Scheme of optical transitions relevant for trion binding. The experimental starting point is the lower left level, with electrostatically introduced carriers in the lower conduction band. The bright optical transitions are those of frequencies $\omega_{\mathrm{BX}}$ and $\omega_T$. The dark exciton transition $\omega_{DX}$ is typically only visible in photoluminescence experiments and not in reflection experiments. b-c)Schematic overview of the b) WSe$_2$ and c) MoSe$_2$ band structure near the K and K' points and the labeling of the trions, composed of the bright exciton and the charges indicated by numbers.}
    \label{fig:bandstructures}
\end{figure}

When comparing to experiment, we focus on the bright peaks at energies $\omega_\mathrm{BX}$ and $\omega_\mathrm{T}$ in Fig.~\ref{fig:bandstructures}a).Here the starting point is a system with electrostatically introduced charge carriers in the lower conduction band.
 In photoluminescence spectra, emission from dark excitons can also be observed, corresponding to the energy $\omega_{\mathrm{DX}}$. In WSe$_2$, the dark exciton peak lies below the trion peaks ($\omega_{\mathrm{DX}}<\omega_T$) \cite{vantuan:2022}. This is because the final states after the transition are different. Hence, the splitting between the bright and dark exciton peaks includes both the difference in exciton binding energies and the conduction band spin–orbit splitting \cite{zhang:2017,robert:2020}.

\textit{Results-} First, we compute the trion energies assuming symmetric trions, neglecting exchange interactions. We use the experimentally fitted Rytova-Keldysh parameters and reduced masses from Ref.~\cite{goryca:2019}, except that we consistently set $\kappa=4.4$ (see \footnote{Compared to Ref.~\cite{goryca:2019}, we we fix $\kappa=4.4$ for both materials instead of the value of $\kappa=4.5$ taken for WSe$_2$ in \cite{goryca:2019}. We change $r_0$ accordingly to keep the exciton binding energy the same.}). For the negatively charged trions in WSe$_2$ [T2 and T3, see Fig.~\ref{fig:bandstructures}b)] we symmetrize the trion equations of motion by using the reduced mass $\bar{\mu}=\frac{2 \mu_1 \mu_2}{\mu_1 + \mu_2}$ \cite{semina:2023} instead of $\mu_1$ and $\mu_2$ [see Eq.~\eqref{eq:kinham}], where $\mu_1$ and $\mu_2$ are the reduced masses of the bright and the dark exciton. The results are shown in the third column of Table \ref{tab:trion_bindings}. 

We find that the theory systematically underestimates the measured results. Previous theoretical works have changed $\kappa=3.3 $ to recover the observed trion binding energies \cite{courtade:2017,hichri:2020}. However, such a large change in $\kappa$ would lead to substantial discrepancies from the experimentally measured Rydberg exciton series \cite{stier:2018,goryca:2019}. Below, we demonstrate that trion binding energies consistent with experiment can be obtained with the experimentally determined $\kappa=4.4$, once the dark-bright exciton splitting and the asymmetry of the trion are accounted for.

\begin{figure}[!b]
    \centering
\includegraphics[width=\columnwidth]{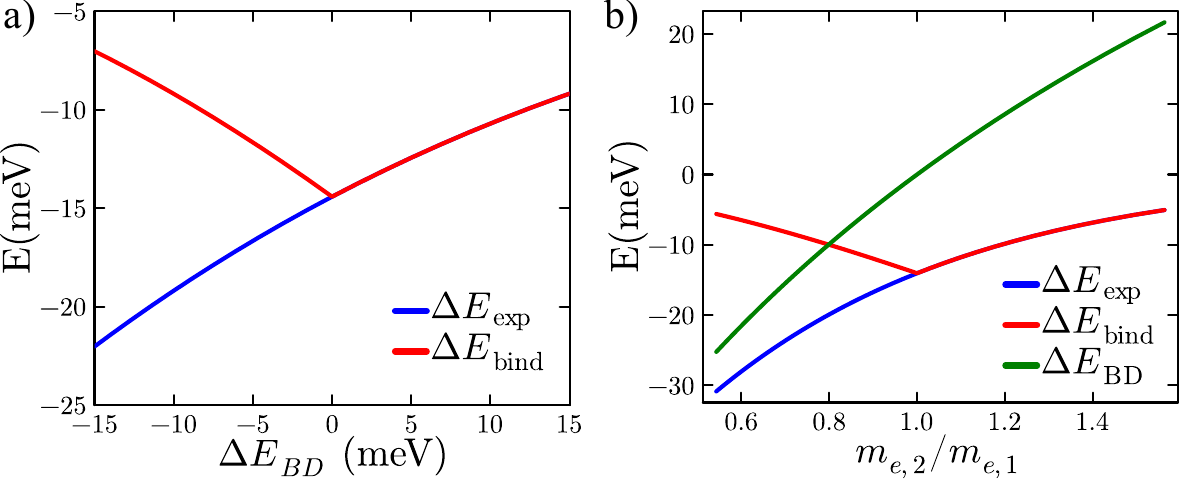}
    \caption{a) The energies $\Delta E_{\mathrm{bind}}$, $\Delta E_{\mathrm{exp}}$ as a function of the splitting between the bright and dark exciton. To vary the splitting, an exchange interaction with varying strength was added to the bright exciton potential. We use the parameters for the hole-trion in WSe$_2$. b) The energies $\Delta E_{\mathrm{bind}}$, $\Delta E_{\mathrm{exp}}$ and $\Delta E_{BD}$ in WSe$_2$, as a function of $m_{e,2}/m_{e,1}$: the ratio between the upper conduction band electron mass $m_{e,2}$ and the lower conduction band electron mass $m_{e,1}$. The value of $m_{e,1}$ is varied, while $m_{e,2}$ is kept fixed. Here we neglect the exchange interactions. }. 
    \label{fig:trionbinding_vsm}
\end{figure}

Fig.~\ref{fig:trionbinding_vsm}a) illustrates the effect of introducing electron-hole exchange interactions. All parameters are fixed to the case of the positively charged trion (T1) in WSe$_2$, and we add a variable-strength exchange interaction to the bright exciton potential. The blue curve shows the trion energy measured with respect to the bright exciton, while the red line shows the trion binding energy. 

We find that the trion binding energy decreases for both attractive and repulsive exchange interactions. This can be understood by considering that in the symmetric configuration the binding between the exciton and the additional electron is the strongest. The small asymmetry observed in the figure arises because adding a net attraction produces a slightly stronger binding than adding a net repulsion.

On the left side of the plot, the trion binding energy is reduced, but the energy splitting between the trion and the bright exciton increases.
This is because the exchange-induced dark–bright exciton splitting $\Delta E_{\mathrm{BD}}$ must be added to the trion binding energy $\Delta E_{\mathrm{bind}}$ to obtain the experimentally observed quantity $\Delta E_{\mathrm{exp}}$. 
We find that the change of $\Delta E_{\mathrm{exp}}$ is approximately half the bright-dark exciton splitting. As shown in Fig.~\ref{fig:trion_scheme}b) , incorporating the theoretically estimated exchange splitting of 11.5 meV brings $\Delta E_{\mathrm{exp}}$ into much better agreement with the experimentally measured value of 21 meV \cite{courtade:2017}.

We next consider the triplet electron trion in WSe$_2$ (T2 in Fig.~\ref{fig:bandstructures}) and analyze the effect of the mass difference between the upper and lower conduction bands. Here, we neglect the exchange interactions. As shown in Fig.~\ref{fig:trionbinding_vsm}b), the trend is similar to that observed in Fig.~\ref{fig:trionbinding_vsm}a): the trion binding energy decreases by approximately half the splitting between the exciton states.

\begin{table}[t!]
    \centering
    \caption{Trion energies (in meV) with respect to the bright exciton in TMDs from experiment and theory.  See Fig.~\ref{fig:bandstructures}b) and c) for the labeling of the trions. Both the symmetric and asymmetric trion scenarios are considered. For the results of the full trion calculations, we break the value of $\Delta E_{\mathrm{exp}}$ down into its two contributions $\Delta E_{\mathrm{BD}}$ and $ \Delta E_{\mathrm{bind}}$.}
    \begin{tabular}{c|c | c | c c c}
    \hline
    \hline
        Trion  &  Exp. & Theory, sym. \footnote{We set $\kappa=4.4$. For WSe$_2$ we use $m_h=0.45$, $m_{e,1}=0.54$, $m_{e,2}=0.36$, $r_0=4.4$nm, and for  MoSe$_2$ we use $m_h=0.6$, $m_e=0.84$, $r_0=3.9$nm.} &  \multicolumn{3}{c}{Theory, full
        \footnote{Same as above, except for $r_0=3.85$nm for WSe$_2$, and $r_0=3.64$nm for MoSe$_2$.}} \\
         & $\Delta E_{\mathrm{exp}}$ & $\Delta E_{\mathrm{exp}}$ & $\Delta E_{\mathrm{exp}}$ & $\Delta E_{\mathrm{BD}}$ & $\Delta E_{\mathrm{bind}}$ \\
        \hline
        WSe$_2$ T1  &  -21 \cite{courtade:2017} & -14.4 &-21.2 & -11.5 & -9.6 \\
        WSe$_2$ T2  &  -29 \cite{courtade:2017} & -22.9 &-26.0 & -19.3 & -6.6 \\
        WSe$_2$ T3  &  -35 \cite{courtade:2017} & -22.9 &-35.0 &-30.7 &-4.2 \\
        MoSe$_2$ T1 &  -24 \cite{kiper:2025} & -19.3 & -24.3 & -9.0 &-15.3 \\
        \hline
        \hline
    \end{tabular}
    \label{tab:trion_bindings}
\end{table}

\begin{figure*}[!t]
    \centering
    \includegraphics[width=\textwidth]{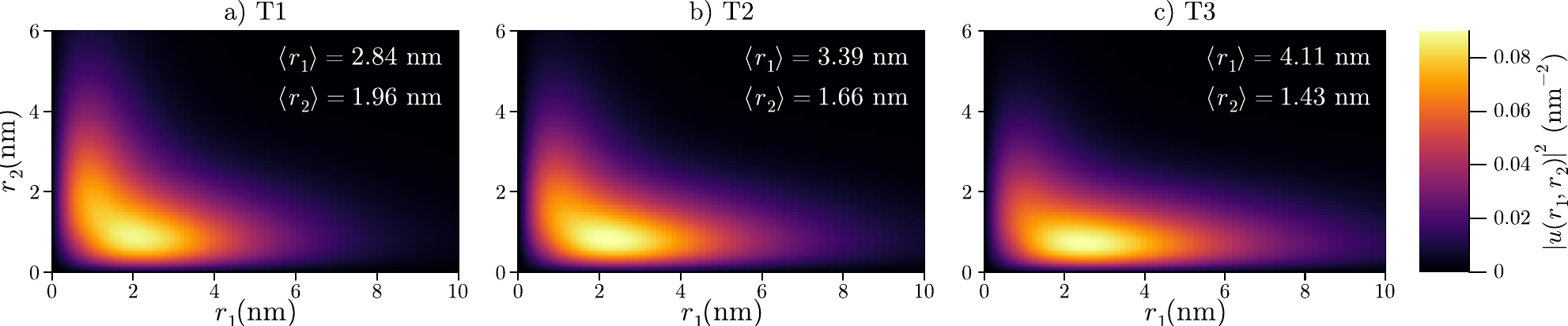}
    \caption{Wavefunctions $|u(r_1,r_2)|^2$ as a function of the electron-hole distances $r_1$ and $r_2$ for the trions a) T1, b) T2 and c) T3 in WSe$_2$ [see Fig.~\ref{fig:bandstructures})].}
    \label{fig:trionwfs}
\end{figure*}

Before turning to quantitative estimates of the asymmetric trion energies, we revisit the experimental calibration of the Rytova–Keldysh screening parameters. The experimentally fitted value of $r_0$ \cite{goryca:2019} was predominantly determined from the 1s-exciton binding energy, but without accounting for the exchange interactions. Assuming the exchange energies calculated in Ref.~\cite{li:2022}, we choose a new $r_0$ that yields the same 1s exciton binding energy. For WSe$_2$ this reduces $r_0$ from $r_0=4.4$ nm to 3.85 nm (for $\kappa=4.4)$. 

Using this updated value of $r_0$ and including the exchange splitting, we obtain a T1 trion energy of 21.2 meV, in remarkable agreement with the experimentally measured value of 21 meV.

The electron trion energies cannot be estimated with the same accuracy, since the electron mass in the lower conduction bands has not been experimentally determined accurately. While theoretical values have been computed,  the experimentally determined reduced mass for the bright exciton deviates by 15–30\% from theory \cite{goryca:2019}, setting the scale for the uncertainty in the theoretically estimated masses.

We reproduce the experimental 35 meV electron-trion (singlet) energy by assuming a 33 \% mass difference \footnote{measured with respect to the lower conduction band electron mass} between the upper and lower conduction band electrons, compared to the theoretically predicted value of approximately 27 \% \cite{kormanyos:2015,li:2022}. For this electron mass and the exchange interaction energy from Ref.~\cite{li:2022}, we obtain a triplet electron-trion energy of 26 meV, compared to the experimental 29 meV.

In MoSe$_2$ the experimental trion binding energy is consistent with an exchange interaction on the order of 9 meV. This implies that the exchange interaction has a similar magnitude as in WSe$_2$, as expected. Furthermore, in Ref.~\cite{robert:2020} an exchange interaction of about 9 meV was estimated for MoS$_2$. Unlike in WSe$_2$, in MoSe$_2$ the dominant contribution to $\Delta E_{\mathrm{exp}}$ arises from the actual trion binding, since this is intrinsically stronger in MoSe$_2$ due to the larger carrier masses.

We expect several mechanisms to give rise to few-meV shifts of the trion energies compared to the results presented here. These include exchange interactions between electrons and holes of opposite spin, exchange interactions between the electrons \cite{dery:2016}, deviations from the Rytova-Keldysh potentials (which may differ for the upper and lower conduction bands), and uncertainties in the electron and hole masses. Experimentally, sample-to-sample variations, local strain, and uncertainties in extracting the trion energy from the optical spectra also contribute at the meV level \cite{kiper:2025,mulkerin:2023}. However, none of these effects are sufficient to account for the large energy difference between the symmetric trion model and the experimentally observed energies, especially in WSe$_2$.

The asymmetric nature of the trions in WSe$_2$ is directly visualized in Fig.~\ref{fig:trionwfs}, which shows the trion wave functions $|u(r_1,r_2)|^2$ as a function of bond lengths $r_1$ and $r_2$, corresponding to the bright and dark excitons, respectively. For the T1-trion [Fig.~\ref{fig:trionwfs}a)], the wave function contains two branches: one where the bright exciton inside the trion is more tightly bound and one where the dark exciton is more tightly bound. In the absence of exchange interactions, these branches would contribute equally. However, the presence of exchange shifts the balance in favor of the branch where the dark exciton is more tightly bound, as reflected in the expectation values $\langle r_1 \rangle = 2.84$ nm and $\langle r_2 \rangle = 1.96$ nm.

For the T3 trion, where $\Delta E_{\mathrm{BD}}$ is the largest, the trion wave function is almost entirely concentrated in the branch where $r_2<r_1$. The bright exciton bond length is stretched, with $\langle r_1 \rangle=4.11$ nm, about twice the typical size assumed for trions. This reflects the small binding energy of only 4.2 meV relative to the dark exciton. As in Table \ref{tab:trion_bindings}, the T2-trion exhibits intermediate behavior between the T1 and T3 cases.

In addition to the optically bright trions, dark trions can also play a key role in the interpretation of the optical spectra in TMDs. In WSe$_2$, with increasing electron density, the trion peaks T2 and T3 vanish, and a new feature interpreted as a hexciton \cite{vantuan:2022} emerges. This excitonic complex consists of a dark trion core, involving the lower conduction band electrons 2 and 3 from Fig.~\ref{fig:bandstructures}, dressed by the holes these electrons leave behind in their respective Fermi Seas. The feature becomes optically bright because of the attachment of an upper conduction band electron as a satellite particle. The disappearance of the T2 and T3 peaks can be attributed to inelastic decay processes with Fermi-sea electrons, which convert the bright trions into dark trions or hexcitons.

We calculate that the binding energy of the dark trion is 10.9 meV with respect to the lowest dark exciton energy, placing its energy 41.6 meV below the bright exciton. This value closely matches the experimental onset of the hexciton feature \cite{vantuan:2022}, supporting the interpretation of its origin. At higher carrier densities, even larger excitonic complexes have been reported \cite{vantuan:2022,dijkstra:2025}.

\textit{Conclusion and outlook-} 
We have studied trions in monolayer WSe$_2$ and MoSe$_2$, demonstrating that exchange interactions and mass differences between the conduction bands lead to strongly asymmetric trion wave functions. This asymmetry plays a crucial role in determining trion binding energies and must be included for quantitative agreement with experimental optical spectra using realistic interaction potentials. In WSe$_2$, we find that the 35 meV splitting between the lowest-energy trion and the bright exciton signatures primarily reflects the bright–dark exciton energy difference, with only 4 meV attributable to true trion binding. Consequently, this trion is roughly twice as spatially extended as commonly assumed.

These insights generalize to other TMDs. In particular, negatively charged trions in WS$_2$ \cite{zipfel:2020} exhibit similar behavior to WSe$_2$. The spectra in MoS$_2$ are more similar to those in MoSe$_2$, with additional complexity likely arising from valley hybridization \cite{klein:2022}.

Our work lays the foundation for a quantitative understanding of exciton-polaron spectra in both monolayer and multilayer TMD systems. A key next step is to incorporate the internal structure of excitons and trions into polaron models \cite{sidler:2017,efimkin:2017,fey:2020,massignan:2025}, which is essential for describing optical responses at higher carrier densities. A quantitative theory based on this framework will enable more detailed interpretation of experimental spectra: the better the few-body aspects of trion binding are understood, the more information can be extracted about both the moiré potential \cite{kiper:2025} and electronic correlations.

\textit{Acknowledgements-} The authors acknowledge useful discussions with Haydn Adlong, Martin Kroner, Mikhail Glazov and Alexey Chernikov. A. C. is supported by an ETH Fellowship. 
A.I. is supported by  SNSF funding under Grant Number 200021-204076. 

\bibliography{references}

\providecommand{\noopsort}[1]{}\providecommand{\singleletter}[1]{#1}%
\begin{thebibliography}{34}%
\makeatletter
\providecommand \@ifxundefined [1]{%
 \@ifx{#1\undefined}
}%
\providecommand \@ifnum [1]{%
 \ifnum #1\expandafter \@firstoftwo
 \else \expandafter \@secondoftwo
 \fi
}%
\providecommand \@ifx [1]{%
 \ifx #1\expandafter \@firstoftwo
 \else \expandafter \@secondoftwo
 \fi
}%
\providecommand \natexlab [1]{#1}%
\providecommand \enquote  [1]{``#1''}%
\providecommand \bibnamefont  [1]{#1}%
\providecommand \bibfnamefont [1]{#1}%
\providecommand \citenamefont [1]{#1}%
\providecommand \href@noop [0]{\@secondoftwo}%
\providecommand \href [0]{\begingroup \@sanitize@url \@href}%
\providecommand \@href[1]{\@@startlink{#1}\@@href}%
\providecommand \@@href[1]{\endgroup#1\@@endlink}%
\providecommand \@sanitize@url [0]{\catcode `\\12\catcode `\$12\catcode
  `\&12\catcode `\#12\catcode `\^12\catcode `\_12\catcode `\%12\relax}%
\providecommand \@@startlink[1]{}%
\providecommand \@@endlink[0]{}%
\providecommand \url  [0]{\begingroup\@sanitize@url \@url }%
\providecommand \@url [1]{\endgroup\@href {#1}{\urlprefix }}%
\providecommand \urlprefix  [0]{URL }%
\providecommand \Eprint [0]{\href }%
\providecommand \doibase [0]{https://doi.org/}%
\providecommand \selectlanguage [0]{\@gobble}%
\providecommand \bibinfo  [0]{\@secondoftwo}%
\providecommand \bibfield  [0]{\@secondoftwo}%
\providecommand \translation [1]{[#1]}%
\providecommand \BibitemOpen [0]{}%
\providecommand \bibitemStop [0]{}%
\providecommand \bibitemNoStop [0]{.\EOS\space}%
\providecommand \EOS [0]{\spacefactor3000\relax}%
\providecommand \BibitemShut  [1]{\csname bibitem#1\endcsname}%
\let\auto@bib@innerbib\@empty
\bibitem [{\citenamefont {Smole{\'n}ski}\ \emph {et~al.}(2021)\citenamefont
  {Smole{\'n}ski}, \citenamefont {Dolgirev}, \citenamefont {Kuhlenkamp},
  \citenamefont {Popert}, \citenamefont {Shimazaki}, \citenamefont {Back},
  \citenamefont {Lu}, \citenamefont {Kroner}, \citenamefont {Watanabe},
  \citenamefont {Taniguchi} \emph {et~al.}}]{smolenski:2021}%
  \BibitemOpen
  \bibfield  {author} {\bibinfo {author} {\bibfnamefont {T.}~\bibnamefont
  {Smole{\'n}ski}}, \bibinfo {author} {\bibfnamefont {P.~E.}\ \bibnamefont
  {Dolgirev}}, \bibinfo {author} {\bibfnamefont {C.}~\bibnamefont
  {Kuhlenkamp}}, \bibinfo {author} {\bibfnamefont {A.}~\bibnamefont {Popert}},
  \bibinfo {author} {\bibfnamefont {Y.}~\bibnamefont {Shimazaki}}, \bibinfo
  {author} {\bibfnamefont {P.}~\bibnamefont {Back}}, \bibinfo {author}
  {\bibfnamefont {X.}~\bibnamefont {Lu}}, \bibinfo {author} {\bibfnamefont
  {M.}~\bibnamefont {Kroner}}, \bibinfo {author} {\bibfnamefont
  {K.}~\bibnamefont {Watanabe}}, \bibinfo {author} {\bibfnamefont
  {T.}~\bibnamefont {Taniguchi}}, \emph {et~al.},\ }\href
  {https://doi.org/10.1038/s41586-021-03590-4} {\bibfield  {journal} {\bibinfo
  {journal} {Nature}\ }\textbf {\bibinfo {volume} {595}},\ \bibinfo {pages}
  {53} (\bibinfo {year} {2021})}\BibitemShut {NoStop}%
\bibitem [{\citenamefont {Cai}\ \emph {et~al.}(2023)\citenamefont {Cai},
  \citenamefont {Anderson}, \citenamefont {Wang}, \citenamefont {Zhang},
  \citenamefont {Liu}, \citenamefont {Holtzmann}, \citenamefont {Zhang},
  \citenamefont {Fan}, \citenamefont {Taniguchi}, \citenamefont {Watanabe}
  \emph {et~al.}}]{cai:2023}%
  \BibitemOpen
  \bibfield  {author} {\bibinfo {author} {\bibfnamefont {J.}~\bibnamefont
  {Cai}}, \bibinfo {author} {\bibfnamefont {E.}~\bibnamefont {Anderson}},
  \bibinfo {author} {\bibfnamefont {C.}~\bibnamefont {Wang}}, \bibinfo {author}
  {\bibfnamefont {X.}~\bibnamefont {Zhang}}, \bibinfo {author} {\bibfnamefont
  {X.}~\bibnamefont {Liu}}, \bibinfo {author} {\bibfnamefont {W.}~\bibnamefont
  {Holtzmann}}, \bibinfo {author} {\bibfnamefont {Y.}~\bibnamefont {Zhang}},
  \bibinfo {author} {\bibfnamefont {F.}~\bibnamefont {Fan}}, \bibinfo {author}
  {\bibfnamefont {T.}~\bibnamefont {Taniguchi}}, \bibinfo {author}
  {\bibfnamefont {K.}~\bibnamefont {Watanabe}}, \emph {et~al.},\ }\href
  {https://doi.org/10.1038/s41586-023-06289-w} {\bibfield  {journal} {\bibinfo
  {journal} {Nature}\ }\textbf {\bibinfo {volume} {622}},\ \bibinfo {pages}
  {63} (\bibinfo {year} {2023})}\BibitemShut {NoStop}%
\bibitem [{\citenamefont {Xu}\ \emph {et~al.}(2023)\citenamefont {Xu},
  \citenamefont {Sun}, \citenamefont {Jia}, \citenamefont {Liu}, \citenamefont
  {Xu}, \citenamefont {Li}, \citenamefont {Gu}, \citenamefont {Watanabe},
  \citenamefont {Taniguchi}, \citenamefont {Tong} \emph {et~al.}}]{xu:2023}%
  \BibitemOpen
  \bibfield  {author} {\bibinfo {author} {\bibfnamefont {F.}~\bibnamefont
  {Xu}}, \bibinfo {author} {\bibfnamefont {Z.}~\bibnamefont {Sun}}, \bibinfo
  {author} {\bibfnamefont {T.}~\bibnamefont {Jia}}, \bibinfo {author}
  {\bibfnamefont {C.}~\bibnamefont {Liu}}, \bibinfo {author} {\bibfnamefont
  {C.}~\bibnamefont {Xu}}, \bibinfo {author} {\bibfnamefont {C.}~\bibnamefont
  {Li}}, \bibinfo {author} {\bibfnamefont {Y.}~\bibnamefont {Gu}}, \bibinfo
  {author} {\bibfnamefont {K.}~\bibnamefont {Watanabe}}, \bibinfo {author}
  {\bibfnamefont {T.}~\bibnamefont {Taniguchi}}, \bibinfo {author}
  {\bibfnamefont {B.}~\bibnamefont {Tong}}, \emph {et~al.},\ }\href
  {https://doi.org/10.1103/PhysRevX.13.031037} {\bibfield  {journal} {\bibinfo
  {journal} {Phys. Rev. X}\ }\textbf {\bibinfo {volume} {13}},\ \bibinfo
  {pages} {031037} (\bibinfo {year} {2023})}\BibitemShut {NoStop}%
\bibitem [{\citenamefont {Zeng}\ \emph {et~al.}(2023)\citenamefont {Zeng},
  \citenamefont {Xia}, \citenamefont {Kang}, \citenamefont {Zhu}, \citenamefont
  {Kn{\"u}ppel}, \citenamefont {Vaswani}, \citenamefont {Watanabe},
  \citenamefont {Taniguchi}, \citenamefont {Mak},\ and\ \citenamefont
  {Shan}}]{zeng:2023}%
  \BibitemOpen
  \bibfield  {author} {\bibinfo {author} {\bibfnamefont {Y.}~\bibnamefont
  {Zeng}}, \bibinfo {author} {\bibfnamefont {Z.}~\bibnamefont {Xia}}, \bibinfo
  {author} {\bibfnamefont {K.}~\bibnamefont {Kang}}, \bibinfo {author}
  {\bibfnamefont {J.}~\bibnamefont {Zhu}}, \bibinfo {author} {\bibfnamefont
  {P.}~\bibnamefont {Kn{\"u}ppel}}, \bibinfo {author} {\bibfnamefont
  {C.}~\bibnamefont {Vaswani}}, \bibinfo {author} {\bibfnamefont
  {K.}~\bibnamefont {Watanabe}}, \bibinfo {author} {\bibfnamefont
  {T.}~\bibnamefont {Taniguchi}}, \bibinfo {author} {\bibfnamefont {K.~F.}\
  \bibnamefont {Mak}},\ and\ \bibinfo {author} {\bibfnamefont {J.}~\bibnamefont
  {Shan}},\ }\href {https://doi.org/10.1038/s41586-023-06452-3} {\bibfield
  {journal} {\bibinfo  {journal} {Nature}\ }\textbf {\bibinfo {volume} {622}},\
  \bibinfo {pages} {69} (\bibinfo {year} {2023})}\BibitemShut {NoStop}%
\bibitem [{\citenamefont {Xia}\ \emph {et~al.}(2025)\citenamefont {Xia},
  \citenamefont {Han}, \citenamefont {Watanabe}, \citenamefont {Taniguchi},
  \citenamefont {Shan},\ and\ \citenamefont {Mak}}]{xia:2025}%
  \BibitemOpen
  \bibfield  {author} {\bibinfo {author} {\bibfnamefont {Y.}~\bibnamefont
  {Xia}}, \bibinfo {author} {\bibfnamefont {Z.}~\bibnamefont {Han}}, \bibinfo
  {author} {\bibfnamefont {K.}~\bibnamefont {Watanabe}}, \bibinfo {author}
  {\bibfnamefont {T.}~\bibnamefont {Taniguchi}}, \bibinfo {author}
  {\bibfnamefont {J.}~\bibnamefont {Shan}},\ and\ \bibinfo {author}
  {\bibfnamefont {K.~F.}\ \bibnamefont {Mak}},\ }\href
  {https://doi.org/10.1038/s41586-024-08116-2} {\bibfield  {journal} {\bibinfo
  {journal} {Nature}\ }\textbf {\bibinfo {volume} {637}},\ \bibinfo {pages}
  {833} (\bibinfo {year} {2025})}\BibitemShut {NoStop}%
\bibitem [{\citenamefont {Guo}\ \emph {et~al.}(2025)\citenamefont {Guo},
  \citenamefont {Pack}, \citenamefont {Swann}, \citenamefont {Holtzman},
  \citenamefont {Cothrine}, \citenamefont {Watanabe}, \citenamefont
  {Taniguchi}, \citenamefont {Mandrus}, \citenamefont {Barmak}, \citenamefont
  {Hone} \emph {et~al.}}]{guo:2025}%
  \BibitemOpen
  \bibfield  {author} {\bibinfo {author} {\bibfnamefont {Y.}~\bibnamefont
  {Guo}}, \bibinfo {author} {\bibfnamefont {J.}~\bibnamefont {Pack}}, \bibinfo
  {author} {\bibfnamefont {J.}~\bibnamefont {Swann}}, \bibinfo {author}
  {\bibfnamefont {L.}~\bibnamefont {Holtzman}}, \bibinfo {author}
  {\bibfnamefont {M.}~\bibnamefont {Cothrine}}, \bibinfo {author}
  {\bibfnamefont {K.}~\bibnamefont {Watanabe}}, \bibinfo {author}
  {\bibfnamefont {T.}~\bibnamefont {Taniguchi}}, \bibinfo {author}
  {\bibfnamefont {D.~G.}\ \bibnamefont {Mandrus}}, \bibinfo {author}
  {\bibfnamefont {K.}~\bibnamefont {Barmak}}, \bibinfo {author} {\bibfnamefont
  {J.}~\bibnamefont {Hone}}, \emph {et~al.},\ }\href
  {https://doi.org/10.1038/s41586-024-08381-1} {\bibfield  {journal} {\bibinfo
  {journal} {Nature}\ }\textbf {\bibinfo {volume} {637}},\ \bibinfo {pages}
  {839} (\bibinfo {year} {2025})}\BibitemShut {NoStop}%
\bibitem [{\citenamefont {Kiper}\ \emph {et~al.}(2025)\citenamefont {Kiper},
  \citenamefont {Adlong}, \citenamefont {Christianen}, \citenamefont {Kroner},
  \citenamefont {Watanabe}, \citenamefont {Taniguchi},\ and\ \citenamefont
  {{\.I}mamo{\u{g}}lu}}]{kiper:2025}%
  \BibitemOpen
  \bibfield  {author} {\bibinfo {author} {\bibfnamefont {N.}~\bibnamefont
  {Kiper}}, \bibinfo {author} {\bibfnamefont {H.~S.}\ \bibnamefont {Adlong}},
  \bibinfo {author} {\bibfnamefont {A.}~\bibnamefont {Christianen}}, \bibinfo
  {author} {\bibfnamefont {M.}~\bibnamefont {Kroner}}, \bibinfo {author}
  {\bibfnamefont {K.}~\bibnamefont {Watanabe}}, \bibinfo {author}
  {\bibfnamefont {T.}~\bibnamefont {Taniguchi}},\ and\ \bibinfo {author}
  {\bibfnamefont {A.}~\bibnamefont {{\.I}mamo{\u{g}}lu}},\ }\href
  {https://doi.org/10.1103/PhysRevX.15.011049} {\bibfield  {journal} {\bibinfo
  {journal} {Phys. Rev. X}\ }\textbf {\bibinfo {volume} {15}},\ \bibinfo
  {pages} {011049} (\bibinfo {year} {2025})}\BibitemShut {NoStop}%
\bibitem [{\citenamefont {Van~Tuan}\ \emph {et~al.}(2022)\citenamefont
  {Van~Tuan}, \citenamefont {Shi}, \citenamefont {Xu}, \citenamefont
  {Crooker},\ and\ \citenamefont {Dery}}]{vantuan:2022}%
  \BibitemOpen
  \bibfield  {author} {\bibinfo {author} {\bibfnamefont {D.}~\bibnamefont
  {Van~Tuan}}, \bibinfo {author} {\bibfnamefont {S.-F.}\ \bibnamefont {Shi}},
  \bibinfo {author} {\bibfnamefont {X.}~\bibnamefont {Xu}}, \bibinfo {author}
  {\bibfnamefont {S.~A.}\ \bibnamefont {Crooker}},\ and\ \bibinfo {author}
  {\bibfnamefont {H.}~\bibnamefont {Dery}},\ }\href
  {https://doi.org/10.1103/PhysRevLett.129.076801} {\bibfield  {journal}
  {\bibinfo  {journal} {Phys. Rev. Lett.}\ }\textbf {\bibinfo {volume} {129}},\
  \bibinfo {pages} {076801} (\bibinfo {year} {2022})}\BibitemShut {NoStop}%
\bibitem [{\citenamefont {Dijkstra}\ \emph {et~al.}(2025)\citenamefont
  {Dijkstra}, \citenamefont {Mhenni}, \citenamefont {Van~Tuan}, \citenamefont
  {{\c{C}}etiner}, \citenamefont {Schur-Wilkens}, \citenamefont {Kim},
  \citenamefont {Steiner}, \citenamefont {Watanabe}, \citenamefont {Taniguchi},
  \citenamefont {Barbone} \emph {et~al.}}]{dijkstra:2025}%
  \BibitemOpen
  \bibfield  {author} {\bibinfo {author} {\bibfnamefont {A.}~\bibnamefont
  {Dijkstra}}, \bibinfo {author} {\bibfnamefont {A.~B.}\ \bibnamefont
  {Mhenni}}, \bibinfo {author} {\bibfnamefont {D.}~\bibnamefont {Van~Tuan}},
  \bibinfo {author} {\bibfnamefont {E.}~\bibnamefont {{\c{C}}etiner}}, \bibinfo
  {author} {\bibfnamefont {M.}~\bibnamefont {Schur-Wilkens}}, \bibinfo {author}
  {\bibfnamefont {J.}~\bibnamefont {Kim}}, \bibinfo {author} {\bibfnamefont
  {L.}~\bibnamefont {Steiner}}, \bibinfo {author} {\bibfnamefont
  {K.}~\bibnamefont {Watanabe}}, \bibinfo {author} {\bibfnamefont
  {T.}~\bibnamefont {Taniguchi}}, \bibinfo {author} {\bibfnamefont
  {M.}~\bibnamefont {Barbone}}, \emph {et~al.},\ }\href@noop {} {\bibfield
  {journal} {\bibinfo  {journal} {arXiv:2505.08923}\ } (\bibinfo {year}
  {2025})}\BibitemShut {NoStop}%
\bibitem [{\citenamefont {Sidler}\ \emph {et~al.}(2017)\citenamefont {Sidler},
  \citenamefont {Back}, \citenamefont {Cotlet}, \citenamefont {Srivastava},
  \citenamefont {Fink}, \citenamefont {Kroner}, \citenamefont {Demler},\ and\
  \citenamefont {Imamoglu}}]{sidler:2017}%
  \BibitemOpen
  \bibfield  {author} {\bibinfo {author} {\bibfnamefont {M.}~\bibnamefont
  {Sidler}}, \bibinfo {author} {\bibfnamefont {P.}~\bibnamefont {Back}},
  \bibinfo {author} {\bibfnamefont {O.}~\bibnamefont {Cotlet}}, \bibinfo
  {author} {\bibfnamefont {A.}~\bibnamefont {Srivastava}}, \bibinfo {author}
  {\bibfnamefont {T.}~\bibnamefont {Fink}}, \bibinfo {author} {\bibfnamefont
  {M.}~\bibnamefont {Kroner}}, \bibinfo {author} {\bibfnamefont
  {E.}~\bibnamefont {Demler}},\ and\ \bibinfo {author} {\bibfnamefont
  {A.}~\bibnamefont {Imamoglu}},\ }\href {https://doi.org/10.1038/nphys3949}
  {\bibfield  {journal} {\bibinfo  {journal} {Nat. Phys.}\ }\textbf {\bibinfo
  {volume} {13}},\ \bibinfo {pages} {255} (\bibinfo {year} {2017})}\BibitemShut
  {NoStop}%
\bibitem [{\citenamefont {Efimkin}\ and\ \citenamefont
  {MacDonald}(2017)}]{efimkin:2017}%
  \BibitemOpen
  \bibfield  {author} {\bibinfo {author} {\bibfnamefont {D.~K.}\ \bibnamefont
  {Efimkin}}\ and\ \bibinfo {author} {\bibfnamefont {A.~H.}\ \bibnamefont
  {MacDonald}},\ }\href {https://doi.org/10.1103/PhysRevB.95.035417} {\bibfield
   {journal} {\bibinfo  {journal} {Phys. Rev. B}\ }\textbf {\bibinfo {volume}
  {95}},\ \bibinfo {pages} {035417} (\bibinfo {year} {2017})}\BibitemShut
  {NoStop}%
\bibitem [{\citenamefont {Courtade}\ \emph {et~al.}(2017)\citenamefont
  {Courtade}, \citenamefont {Semina}, \citenamefont {Manca}, \citenamefont
  {Glazov}, \citenamefont {Robert}, \citenamefont {Cadiz}, \citenamefont
  {Wang}, \citenamefont {Taniguchi}, \citenamefont {Watanabe}, \citenamefont
  {Pierre} \emph {et~al.}}]{courtade:2017}%
  \BibitemOpen
  \bibfield  {author} {\bibinfo {author} {\bibfnamefont {E.}~\bibnamefont
  {Courtade}}, \bibinfo {author} {\bibfnamefont {M.}~\bibnamefont {Semina}},
  \bibinfo {author} {\bibfnamefont {M.}~\bibnamefont {Manca}}, \bibinfo
  {author} {\bibfnamefont {M.}~\bibnamefont {Glazov}}, \bibinfo {author}
  {\bibfnamefont {C.}~\bibnamefont {Robert}}, \bibinfo {author} {\bibfnamefont
  {F.}~\bibnamefont {Cadiz}}, \bibinfo {author} {\bibfnamefont
  {G.}~\bibnamefont {Wang}}, \bibinfo {author} {\bibfnamefont {T.}~\bibnamefont
  {Taniguchi}}, \bibinfo {author} {\bibfnamefont {K.}~\bibnamefont {Watanabe}},
  \bibinfo {author} {\bibfnamefont {M.}~\bibnamefont {Pierre}}, \emph
  {et~al.},\ }\href {https://doi.org/10.1103/PhysRevB.96.085302} {\bibfield
  {journal} {\bibinfo  {journal} {Phys. Rev. B}\ }\textbf {\bibinfo {volume}
  {96}},\ \bibinfo {pages} {085302} (\bibinfo {year} {2017})}\BibitemShut
  {NoStop}%
\bibitem [{\citenamefont {Glazov}(2020)}]{glazov:2020}%
  \BibitemOpen
  \bibfield  {author} {\bibinfo {author} {\bibfnamefont {M.~M.}\ \bibnamefont
  {Glazov}},\ }\bibfield  {journal} {\bibinfo  {journal} {J. Chem. Phys.}\
  }\textbf {\bibinfo {volume} {153}},\ \href
  {https://doi.org/10.1063/5.0012475} {10.1063/5.0012475} (\bibinfo {year}
  {2020})\BibitemShut {NoStop}%
\bibitem [{\citenamefont {Stier}\ \emph {et~al.}(2018)\citenamefont {Stier},
  \citenamefont {Wilson}, \citenamefont {Velizhanin}, \citenamefont {Kono},
  \citenamefont {Xu},\ and\ \citenamefont {Crooker}}]{stier:2018}%
  \BibitemOpen
  \bibfield  {author} {\bibinfo {author} {\bibfnamefont {A.~V.}\ \bibnamefont
  {Stier}}, \bibinfo {author} {\bibfnamefont {N.~P.}\ \bibnamefont {Wilson}},
  \bibinfo {author} {\bibfnamefont {K.~A.}\ \bibnamefont {Velizhanin}},
  \bibinfo {author} {\bibfnamefont {J.}~\bibnamefont {Kono}}, \bibinfo {author}
  {\bibfnamefont {X.}~\bibnamefont {Xu}},\ and\ \bibinfo {author}
  {\bibfnamefont {S.~A.}\ \bibnamefont {Crooker}},\ }\href
  {https://doi.org/10.1103/PhysRevLett.120.057405} {\bibfield  {journal}
  {\bibinfo  {journal} {Phys. Rev. Lett.}\ }\textbf {\bibinfo {volume} {120}},\
  \bibinfo {pages} {057405} (\bibinfo {year} {2018})}\BibitemShut {NoStop}%
\bibitem [{\citenamefont {Goryca}\ \emph {et~al.}(2019)\citenamefont {Goryca},
  \citenamefont {Li}, \citenamefont {Stier}, \citenamefont {Taniguchi},
  \citenamefont {Watanabe}, \citenamefont {Courtade}, \citenamefont {Shree},
  \citenamefont {Robert}, \citenamefont {Urbaszek}, \citenamefont {Marie},\
  and\ \citenamefont {Crooker}}]{goryca:2019}%
  \BibitemOpen
  \bibfield  {author} {\bibinfo {author} {\bibfnamefont {M.}~\bibnamefont
  {Goryca}}, \bibinfo {author} {\bibfnamefont {J.}~\bibnamefont {Li}}, \bibinfo
  {author} {\bibfnamefont {A.~V.}\ \bibnamefont {Stier}}, \bibinfo {author}
  {\bibfnamefont {T.}~\bibnamefont {Taniguchi}}, \bibinfo {author}
  {\bibfnamefont {K.}~\bibnamefont {Watanabe}}, \bibinfo {author}
  {\bibfnamefont {E.}~\bibnamefont {Courtade}}, \bibinfo {author}
  {\bibfnamefont {S.}~\bibnamefont {Shree}}, \bibinfo {author} {\bibfnamefont
  {C.}~\bibnamefont {Robert}}, \bibinfo {author} {\bibfnamefont
  {B.}~\bibnamefont {Urbaszek}}, \bibinfo {author} {\bibfnamefont
  {X.}~\bibnamefont {Marie}},\ and\ \bibinfo {author} {\bibfnamefont {S.~A.}\
  \bibnamefont {Crooker}},\ }\href {https://doi.org/10.1038/s41467-019-12180-y}
  {\bibfield  {journal} {\bibinfo  {journal} {Nat. Comm.}\ }\textbf {\bibinfo
  {volume} {10}},\ \bibinfo {pages} {4172} (\bibinfo {year}
  {2019})}\BibitemShut {NoStop}%
\bibitem [{\citenamefont {Li}\ \emph {et~al.}(2022)\citenamefont {Li},
  \citenamefont {Robert}, \citenamefont {Van~Tuan}, \citenamefont {Ren},
  \citenamefont {Yang}, \citenamefont {Marie},\ and\ \citenamefont
  {Dery}}]{li:2022}%
  \BibitemOpen
  \bibfield  {author} {\bibinfo {author} {\bibfnamefont {P.}~\bibnamefont
  {Li}}, \bibinfo {author} {\bibfnamefont {C.}~\bibnamefont {Robert}}, \bibinfo
  {author} {\bibfnamefont {D.}~\bibnamefont {Van~Tuan}}, \bibinfo {author}
  {\bibfnamefont {L.}~\bibnamefont {Ren}}, \bibinfo {author} {\bibfnamefont
  {M.}~\bibnamefont {Yang}}, \bibinfo {author} {\bibfnamefont {X.}~\bibnamefont
  {Marie}},\ and\ \bibinfo {author} {\bibfnamefont {H.}~\bibnamefont {Dery}},\
  }\href {https://doi.org/10.1103/PhysRevB.106.085414} {\bibfield  {journal}
  {\bibinfo  {journal} {Phys. Rev. B}\ }\textbf {\bibinfo {volume} {106}},\
  \bibinfo {pages} {085414} (\bibinfo {year} {2022})}\BibitemShut {NoStop}%
\bibitem [{\citenamefont {Zhang}\ \emph {et~al.}(2017)\citenamefont {Zhang},
  \citenamefont {Cao}, \citenamefont {Lu}, \citenamefont {Lin}, \citenamefont
  {Zhang}, \citenamefont {Wang}, \citenamefont {Li}, \citenamefont {Hone},
  \citenamefont {Robinson}, \citenamefont {Smirnov} \emph
  {et~al.}}]{zhang:2017}%
  \BibitemOpen
  \bibfield  {author} {\bibinfo {author} {\bibfnamefont {X.-X.}\ \bibnamefont
  {Zhang}}, \bibinfo {author} {\bibfnamefont {T.}~\bibnamefont {Cao}}, \bibinfo
  {author} {\bibfnamefont {Z.}~\bibnamefont {Lu}}, \bibinfo {author}
  {\bibfnamefont {Y.-C.}\ \bibnamefont {Lin}}, \bibinfo {author} {\bibfnamefont
  {F.}~\bibnamefont {Zhang}}, \bibinfo {author} {\bibfnamefont
  {Y.}~\bibnamefont {Wang}}, \bibinfo {author} {\bibfnamefont {Z.}~\bibnamefont
  {Li}}, \bibinfo {author} {\bibfnamefont {J.~C.}\ \bibnamefont {Hone}},
  \bibinfo {author} {\bibfnamefont {J.~A.}\ \bibnamefont {Robinson}}, \bibinfo
  {author} {\bibfnamefont {D.}~\bibnamefont {Smirnov}}, \emph {et~al.},\ }\href
  {https://doi.org/10.1038/nnano.2017.105} {\bibfield  {journal} {\bibinfo
  {journal} {Nat. Nanotechnol.}\ }\textbf {\bibinfo {volume} {12}},\ \bibinfo
  {pages} {883} (\bibinfo {year} {2017})}\BibitemShut {NoStop}%
\bibitem [{\citenamefont {Robert}\ \emph {et~al.}(2020)\citenamefont {Robert},
  \citenamefont {Han}, \citenamefont {Kapuscinski}, \citenamefont {Delhomme},
  \citenamefont {Faugeras}, \citenamefont {Amand}, \citenamefont {Molas},
  \citenamefont {Bartos}, \citenamefont {Watanabe}, \citenamefont {Taniguchi}
  \emph {et~al.}}]{robert:2020}%
  \BibitemOpen
  \bibfield  {author} {\bibinfo {author} {\bibfnamefont {C.}~\bibnamefont
  {Robert}}, \bibinfo {author} {\bibfnamefont {B.}~\bibnamefont {Han}},
  \bibinfo {author} {\bibfnamefont {P.}~\bibnamefont {Kapuscinski}}, \bibinfo
  {author} {\bibfnamefont {A.}~\bibnamefont {Delhomme}}, \bibinfo {author}
  {\bibfnamefont {C.}~\bibnamefont {Faugeras}}, \bibinfo {author}
  {\bibfnamefont {T.}~\bibnamefont {Amand}}, \bibinfo {author} {\bibfnamefont
  {M.~R.}\ \bibnamefont {Molas}}, \bibinfo {author} {\bibfnamefont
  {M.}~\bibnamefont {Bartos}}, \bibinfo {author} {\bibfnamefont
  {K.}~\bibnamefont {Watanabe}}, \bibinfo {author} {\bibfnamefont
  {T.}~\bibnamefont {Taniguchi}}, \emph {et~al.},\ }\href
  {https://doi.org/10.1038/s41467-020-17608-4} {\bibfield  {journal} {\bibinfo
  {journal} {Nat. Comm.}\ }\textbf {\bibinfo {volume} {11}},\ \bibinfo {pages}
  {4037} (\bibinfo {year} {2020})}\BibitemShut {NoStop}%
\bibitem [{\citenamefont {Korm{\'a}nyos}\ \emph {et~al.}(2015)\citenamefont
  {Korm{\'a}nyos}, \citenamefont {Burkard}, \citenamefont {Gmitra},
  \citenamefont {Fabian}, \citenamefont {Z{\'o}lyomi}, \citenamefont
  {Drummond},\ and\ \citenamefont {Fal’ko}}]{kormanyos:2015}%
  \BibitemOpen
  \bibfield  {author} {\bibinfo {author} {\bibfnamefont {A.}~\bibnamefont
  {Korm{\'a}nyos}}, \bibinfo {author} {\bibfnamefont {G.}~\bibnamefont
  {Burkard}}, \bibinfo {author} {\bibfnamefont {M.}~\bibnamefont {Gmitra}},
  \bibinfo {author} {\bibfnamefont {J.}~\bibnamefont {Fabian}}, \bibinfo
  {author} {\bibfnamefont {V.}~\bibnamefont {Z{\'o}lyomi}}, \bibinfo {author}
  {\bibfnamefont {N.~D.}\ \bibnamefont {Drummond}},\ and\ \bibinfo {author}
  {\bibfnamefont {V.}~\bibnamefont {Fal’ko}},\ }\href
  {https://doi.org/10.1088/2053-1583/2/2/022001} {\bibfield  {journal}
  {\bibinfo  {journal} {2D Materials}\ }\textbf {\bibinfo {volume} {2}},\
  \bibinfo {pages} {022001} (\bibinfo {year} {2015})}\BibitemShut {NoStop}%
\bibitem [{\citenamefont {Fey}\ \emph {et~al.}(2020)\citenamefont {Fey},
  \citenamefont {Schmelcher}, \citenamefont {Imamoglu},\ and\ \citenamefont
  {Schmidt}}]{fey:2020}%
  \BibitemOpen
  \bibfield  {author} {\bibinfo {author} {\bibfnamefont {C.}~\bibnamefont
  {Fey}}, \bibinfo {author} {\bibfnamefont {P.}~\bibnamefont {Schmelcher}},
  \bibinfo {author} {\bibfnamefont {A.}~\bibnamefont {Imamoglu}},\ and\
  \bibinfo {author} {\bibfnamefont {R.}~\bibnamefont {Schmidt}},\ }\href
  {https://doi.org/10.1103/PhysRevB.101.195417} {\bibfield  {journal} {\bibinfo
   {journal} {Phys. Rev. B}\ }\textbf {\bibinfo {volume} {101}},\ \bibinfo
  {pages} {195417} (\bibinfo {year} {2020})}\BibitemShut {NoStop}%
\bibitem [{\citenamefont {Rytova}(1967)}]{rytova:1967}%
  \BibitemOpen
  \bibfield  {author} {\bibinfo {author} {\bibfnamefont {N.}~\bibnamefont
  {Rytova}},\ }\href@noop {} {\bibfield  {journal} {\bibinfo  {journal} {Proc.
  MSU Phys. Astron.}\ }\textbf {\bibinfo {volume} {3}},\ \bibinfo {pages} {30}
  (\bibinfo {year} {1967})}\BibitemShut {NoStop}%
\bibitem [{\citenamefont {Keldysh}(1979)}]{keldysh:1979}%
  \BibitemOpen
  \bibfield  {author} {\bibinfo {author} {\bibfnamefont {L.}~\bibnamefont
  {Keldysh}},\ }\href@noop {} {\bibfield  {journal} {\bibinfo  {journal}
  {Pis’ma Zh. Eksp. Teor. Fiz.}\ }\textbf {\bibinfo {volume} {29}},\ \bibinfo
  {pages} {716} (\bibinfo {year} {1979})}\BibitemShut {NoStop}%
\bibitem [{\citenamefont {Srivastava}\ and\ \citenamefont
  {Imamo\ifmmode~\breve{g}\else \u{g}\fi{}lu}(2015)}]{srivastava:2015}%
  \BibitemOpen
  \bibfield  {author} {\bibinfo {author} {\bibfnamefont {A.}~\bibnamefont
  {Srivastava}}\ and\ \bibinfo {author} {\bibfnamefont {A.~m.~c.}\ \bibnamefont
  {Imamo\ifmmode~\breve{g}\else \u{g}\fi{}lu}},\ }\href
  {https://doi.org/10.1103/PhysRevLett.115.166802} {\bibfield  {journal}
  {\bibinfo  {journal} {Phys. Rev. Lett.}\ }\textbf {\bibinfo {volume} {115}},\
  \bibinfo {pages} {166802} (\bibinfo {year} {2015})}\BibitemShut {NoStop}%
\bibitem [{\citenamefont {Liu}\ \emph {et~al.}(2013)\citenamefont {Liu},
  \citenamefont {Shan}, \citenamefont {Yao}, \citenamefont {Yao},\ and\
  \citenamefont {Xiao}}]{liu:2013}%
  \BibitemOpen
  \bibfield  {author} {\bibinfo {author} {\bibfnamefont {G.-B.}\ \bibnamefont
  {Liu}}, \bibinfo {author} {\bibfnamefont {W.-Y.}\ \bibnamefont {Shan}},
  \bibinfo {author} {\bibfnamefont {Y.}~\bibnamefont {Yao}}, \bibinfo {author}
  {\bibfnamefont {W.}~\bibnamefont {Yao}},\ and\ \bibinfo {author}
  {\bibfnamefont {D.}~\bibnamefont {Xiao}},\ }\href
  {https://doi.org/10.1103/PhysRevB.88.085433} {\bibfield  {journal} {\bibinfo
  {journal} {Phys. Rev. B}\ }\textbf {\bibinfo {volume} {88}},\ \bibinfo
  {pages} {085433} (\bibinfo {year} {2013})}\BibitemShut {NoStop}%
\bibitem [{\citenamefont {Dery}(2016)}]{dery:2016}%
  \BibitemOpen
  \bibfield  {author} {\bibinfo {author} {\bibfnamefont {H.}~\bibnamefont
  {Dery}},\ }\href {https://doi.org/10.1103/PhysRevB.94.075421} {\bibfield
  {journal} {\bibinfo  {journal} {Phys. Rev. B}\ }\textbf {\bibinfo {volume}
  {94}},\ \bibinfo {pages} {075421} (\bibinfo {year} {2016})}\BibitemShut
  {NoStop}%
\bibitem [{Note1()}]{Note1}%
  \BibitemOpen
  \bibinfo {note} {Compared to Ref.~\cite {goryca:2019}, we we fix $\kappa
  =4.4$ for both materials instead of the value of $\kappa =4.5$ taken for
  WSe$_2$ in \cite {goryca:2019}. We change $r_0$ accordingly to keep the
  exciton binding energy the same.}\BibitemShut {Stop}%
\bibitem [{\citenamefont {Semina}\ \emph {et~al.}(2023)\citenamefont {Semina},
  \citenamefont {Mamedov},\ and\ \citenamefont {Glazov}}]{semina:2023}%
  \BibitemOpen
  \bibfield  {author} {\bibinfo {author} {\bibfnamefont {M.~A.}\ \bibnamefont
  {Semina}}, \bibinfo {author} {\bibfnamefont {J.~V.}\ \bibnamefont
  {Mamedov}},\ and\ \bibinfo {author} {\bibfnamefont {M.~M.}\ \bibnamefont
  {Glazov}},\ }\href {https://doi.org/10.1093/oxfmat/itad004} {\bibfield
  {journal} {\bibinfo  {journal} {Oxf. Open Mater. Sci.}\ }\textbf {\bibinfo
  {volume} {3}},\ \bibinfo {pages} {itad004} (\bibinfo {year}
  {2023})}\BibitemShut {NoStop}%
\bibitem [{\citenamefont {Hichri}\ and\ \citenamefont
  {Jaziri}(2020)}]{hichri:2020}%
  \BibitemOpen
  \bibfield  {author} {\bibinfo {author} {\bibfnamefont {A.}~\bibnamefont
  {Hichri}}\ and\ \bibinfo {author} {\bibfnamefont {S.}~\bibnamefont
  {Jaziri}},\ }\href {https://doi.org/10.1103/PhysRevB.102.085407} {\bibfield
  {journal} {\bibinfo  {journal} {Phys. Rev. B}\ }\textbf {\bibinfo {volume}
  {102}},\ \bibinfo {pages} {085407} (\bibinfo {year} {2020})}\BibitemShut
  {NoStop}%
\bibitem [{Note2()}]{Note2}%
  \BibitemOpen
  \bibinfo {note} {Measured with respect to the lower conduction band electron
  mass}\BibitemShut {NoStop}%
\bibitem [{\citenamefont {Mulkerin}\ \emph {et~al.}(2023)\citenamefont
  {Mulkerin}, \citenamefont {Tiene}, \citenamefont {Marchetti}, \citenamefont
  {Parish},\ and\ \citenamefont {Levinsen}}]{mulkerin:2023}%
  \BibitemOpen
  \bibfield  {author} {\bibinfo {author} {\bibfnamefont {B.~C.}\ \bibnamefont
  {Mulkerin}}, \bibinfo {author} {\bibfnamefont {A.}~\bibnamefont {Tiene}},
  \bibinfo {author} {\bibfnamefont {F.~M.}\ \bibnamefont {Marchetti}}, \bibinfo
  {author} {\bibfnamefont {M.~M.}\ \bibnamefont {Parish}},\ and\ \bibinfo
  {author} {\bibfnamefont {J.}~\bibnamefont {Levinsen}},\ }\href
  {https://doi.org/10.1103/PhysRevLett.131.106901} {\bibfield  {journal}
  {\bibinfo  {journal} {Phys. Rev. Lett.}\ }\textbf {\bibinfo {volume} {131}},\
  \bibinfo {pages} {106901} (\bibinfo {year} {2023})}\BibitemShut {NoStop}%
\bibitem [{\citenamefont {Zipfel}\ \emph {et~al.}(2020)\citenamefont {Zipfel},
  \citenamefont {Wagner}, \citenamefont {Ziegler}, \citenamefont {Taniguchi},
  \citenamefont {Watanabe}, \citenamefont {Semina},\ and\ \citenamefont
  {Chernikov}}]{zipfel:2020}%
  \BibitemOpen
  \bibfield  {author} {\bibinfo {author} {\bibfnamefont {J.}~\bibnamefont
  {Zipfel}}, \bibinfo {author} {\bibfnamefont {K.}~\bibnamefont {Wagner}},
  \bibinfo {author} {\bibfnamefont {J.~D.}\ \bibnamefont {Ziegler}}, \bibinfo
  {author} {\bibfnamefont {T.}~\bibnamefont {Taniguchi}}, \bibinfo {author}
  {\bibfnamefont {K.}~\bibnamefont {Watanabe}}, \bibinfo {author}
  {\bibfnamefont {M.~A.}\ \bibnamefont {Semina}},\ and\ \bibinfo {author}
  {\bibfnamefont {A.}~\bibnamefont {Chernikov}},\ }\bibfield  {journal}
  {\bibinfo  {journal} {J. Chem. Phys.}\ }\textbf {\bibinfo {volume} {153}},\
  \href {https://doi.org/10.1063/5.0012721} {10.1063/5.0012721} (\bibinfo
  {year} {2020})\BibitemShut {NoStop}%
\bibitem [{\citenamefont {Klein}\ \emph {et~al.}(2022)\citenamefont {Klein},
  \citenamefont {Florian}, \citenamefont {H\"otger}, \citenamefont {Steinhoff},
  \citenamefont {Delhomme}, \citenamefont {Taniguchi}, \citenamefont
  {Watanabe}, \citenamefont {Jahnke}, \citenamefont {Holleitner}, \citenamefont
  {Potemski}, \citenamefont {Faugeras}, \citenamefont {Stier},\ and\
  \citenamefont {Finley}}]{klein:2022}%
  \BibitemOpen
  \bibfield  {author} {\bibinfo {author} {\bibfnamefont {J.}~\bibnamefont
  {Klein}}, \bibinfo {author} {\bibfnamefont {M.}~\bibnamefont {Florian}},
  \bibinfo {author} {\bibfnamefont {A.}~\bibnamefont {H\"otger}}, \bibinfo
  {author} {\bibfnamefont {A.}~\bibnamefont {Steinhoff}}, \bibinfo {author}
  {\bibfnamefont {A.}~\bibnamefont {Delhomme}}, \bibinfo {author}
  {\bibfnamefont {T.}~\bibnamefont {Taniguchi}}, \bibinfo {author}
  {\bibfnamefont {K.}~\bibnamefont {Watanabe}}, \bibinfo {author}
  {\bibfnamefont {F.}~\bibnamefont {Jahnke}}, \bibinfo {author} {\bibfnamefont
  {A.~W.}\ \bibnamefont {Holleitner}}, \bibinfo {author} {\bibfnamefont
  {M.}~\bibnamefont {Potemski}}, \bibinfo {author} {\bibfnamefont
  {C.}~\bibnamefont {Faugeras}}, \bibinfo {author} {\bibfnamefont {A.~V.}\
  \bibnamefont {Stier}},\ and\ \bibinfo {author} {\bibfnamefont {J.~J.}\
  \bibnamefont {Finley}},\ }\href
  {https://doi.org/10.1103/PhysRevB.105.L041302} {\bibfield  {journal}
  {\bibinfo  {journal} {Phys. Rev. B}\ }\textbf {\bibinfo {volume} {105}},\
  \bibinfo {pages} {L041302} (\bibinfo {year} {2022})}\BibitemShut {NoStop}%
\bibitem [{\citenamefont {Massignan}\ \emph {et~al.}(2025)\citenamefont
  {Massignan}, \citenamefont {Schmidt}, \citenamefont {Astrakharchik},
  \citenamefont {{\.I}mamoglu}, \citenamefont {Zwierlein}, \citenamefont
  {Arlt},\ and\ \citenamefont {Bruun}}]{massignan:2025}%
  \BibitemOpen
  \bibfield  {author} {\bibinfo {author} {\bibfnamefont {P.}~\bibnamefont
  {Massignan}}, \bibinfo {author} {\bibfnamefont {R.}~\bibnamefont {Schmidt}},
  \bibinfo {author} {\bibfnamefont {G.~E.}\ \bibnamefont {Astrakharchik}},
  \bibinfo {author} {\bibfnamefont {A.}~\bibnamefont {{\.I}mamoglu}}, \bibinfo
  {author} {\bibfnamefont {M.}~\bibnamefont {Zwierlein}}, \bibinfo {author}
  {\bibfnamefont {J.~J.}\ \bibnamefont {Arlt}},\ and\ \bibinfo {author}
  {\bibfnamefont {G.~M.}\ \bibnamefont {Bruun}},\ }\href@noop {} {\bibfield
  {journal} {\bibinfo  {journal} {arXiv:2501.09618}\ } (\bibinfo {year}
  {2025})}\BibitemShut {NoStop}%
\bibitem [{\citenamefont {Al-Marzoug}\ \emph {et~al.}(2011)\citenamefont
  {Al-Marzoug}, \citenamefont {Bahlouli},\ and\ \citenamefont
  {Abdelmonem}}]{almarzoug:2011}%
  \BibitemOpen
  \bibfield  {author} {\bibinfo {author} {\bibfnamefont {S.}~\bibnamefont
  {Al-Marzoug}}, \bibinfo {author} {\bibfnamefont {H.}~\bibnamefont
  {Bahlouli}},\ and\ \bibinfo {author} {\bibfnamefont {M.}~\bibnamefont
  {Abdelmonem}},\ }\href@noop {} {\bibfield  {journal} {\bibinfo  {journal}
  {arXiv:1110.0958}\ } (\bibinfo {year} {2011})}\BibitemShut {NoStop}%
\end{thebibliography}%

\onecolumngrid
\section{Supplemental Material} 

For the trion binding calculations, we follow Ref.~\cite{fey:2020}. In brief: we set $\hbar=1$ and assume a quadratic dispersion of the electrons and holes, leading to a Hamiltonian
\begin{equation}
\mathcal{H}=-\sum_{i=1,2,3}\frac{1}{2 m_i} \nabla_{\bm{R}_i}^2 + V_{12}(\bm{R_1}-\bm{R_2})+V_{13}(\bm{R_1}-\bm{R_3})+V_{23}(\bm{R_2}-\bm{R_3}),
\end{equation}
where $\bm{R_i}$ denotes the two-dimensional position of the particle $i$. We define the electron-hole distances $\bm{r_1}=\bm{R_1}-\bm{R_3}$ and $\bm{r_2}=\bm{R_2}-\bm{R_3}$. Then, in the center-of-mass frame, the Hamiltonian becomes
\begin{equation}
\mathcal{H}=-\frac{1}{2 \mu_1} \nabla_{\bm{r}_1}^2 -\frac{1}{2 \mu_2} \nabla_{\bm{r}_2}^2 -\frac{1}{m_3} \nabla_{\bm{r_1}} \cdot \nabla_{\bm{r_2}} + V_{12}(\bm{r_1}-\bm{r_2})+V_{13}(\bm{r_1})+V_{23}(\bm{r_2}),
\end{equation}
where $\mu_1=\frac{m_1 m_3}{m_1+m_3}$ and $\mu_2=\frac{m_2m_3}{m_2+m_3}$.

We now write our wave function as 
\begin{equation}
    \psi_T(r_1,r_2,\theta, \alpha)=\frac{u(r_1,r_2,\theta)}{\sqrt{2\pi r_1 r_2}} \exp(i m \alpha),
\end{equation}
and set $m=0$. Here, $\theta$ is the angle between $\vec{r}_1$ and $\vec{r}_2$, and $\alpha$ is the angle specifying the overall orientation of the trion.

In this case, the kinetic part of the Hamiltonian operator acting on $u(r_1,r_2,\theta)$ is given by \cite{fey:2020}
\begin{multline}\label{eq:kinham}
    \mathcal{H}_{kin}= - \frac{1}{2\mu_1} \left(\frac{\partial^2}{\partial r_1^2} + \frac{1}{4 r_1^2} + \frac{1}{r_1^2} \frac{\partial^2}{\partial \theta^2} \right)- \frac{1}{2\mu_2} \left(\frac{\partial^2}{\partial r_2^2} + \frac{1}{4 r_2^2} + \frac{1}{r_2^2} \frac{\partial^2}{\partial \theta^2} \right) \\
    -\frac{1}{m_3}\left[\frac{1}{r_1 r_2} \left(\frac{\cos \theta}{4}-\frac{\partial}{\partial\theta} \cos \theta \frac{\partial}{\partial\theta}\right) + \cos \theta \frac{\partial^2}{\partial r_1 \partial r_2} -\frac{1}{2} \left( \frac{\partial}{r_1 \partial r_2 }+\frac{\partial}{r_2 \partial r_1 } \right) \left( \sin \theta \frac{\partial}{\partial \theta} + \frac{\partial}{\partial \theta} \sin \theta \right)\right].
\end{multline}
We variationally parameterize our wave function as
\begin{equation} \label{eq:varwf1}
u(r_1,r_2,\theta) = \frac{1}{\sqrt{2 \pi}} \sum_{i,j,l} c_l \phi_{l,i}(r_1) \phi_{l,j}(r_2) \cos(l \theta),
\end{equation}
where $c_0=1$ and $c_{l>0}=\sqrt{2}$. Due to the mirror symmetry, the even and odd functions in $\theta$ decouple, and the ground-state trion lives in the even sector. We have chosen basis functions $\phi$ in the form of 2D-hydrogenic wave functions
\begin{equation} \label{eq:varwf2}
\phi_{l,i}=(\frac{r}{\lambda})^{l+1/2} L_{i}^{2l}(\frac{r}{\lambda})\exp(-\frac{r}{2\lambda}) \sqrt{\frac{\Gamma(i+1)}{\lambda \Gamma(i+2l+1) }}.
\end{equation}
Here. $L_{i}^{2l}(\frac{r}{\lambda})$ is the generalized Laguerre polynomial. With these basis functions the overlap matrix and the Hamiltonian terms in the first line of Eq.~\eqref{eq:kinham} are tridiagonal matrices, see 
Ref.~\cite{almarzoug:2011} (with the inverse definition of $\lambda$). The value of $\lambda$ can be optimized to obtain good convergence with a small number of basis functions (between 8 and 20 per value of $l$). We compute the matrix elements of the potentials numerically. 
\end{document}